\documentclass[conference]{IEEEtran}

\usepackage[letterpaper, left=0.77in, right=0.77in, bottom=1in, top=0.75in]{geometry}
\usepackage{graphicx}
\usepackage{color}

\begin{document}
%
% paper title
% Titles are generally capitalized except for words such as a, an, and, as,
% at, but, by, for, in, nor, of, on, or, the, to and up, which are usually
% not capitalized unless they are the first or last word of the title.
% Linebreaks \\ can be used within to get better formatting as desired.
% Do not put math or special symbols in the title.
\title{Towards Multi-container Deployment\\ on IoT Gateways}

% author names and affiliations
% use a multiple column layout for up to three different
% affiliations
\author{\IEEEauthorblockN{Koustabh Dolui}
\IEEEauthorblockA{imec-Distrinet\\
Katholieke Universiteit Leuven\\
Email: koustabh.dolui@kuleuven.be\\
ORCID: 0000-0002-2690-4038}
\and
\IEEEauthorblockN{Csaba Kiraly}
\IEEEauthorblockA{OpenIoT Research Unit\\
FBK CREATE-NET\\
	Email: kiraly@fbk.eu\\
    ORCID: 0000-0002-6839-5024}
}

% conference papers do not typically use \thanks and this command
% is locked out in conference mode. If really needed, such as for
% the acknowledgment of grants, issue a \IEEEoverridecommandlockouts
% after \documentclass

% for over three affiliations, or if they all won't fit within the width
% of the page, use this alternative format:
% 
%\author{\IEEEauthorblockN{Michael Shell\IEEEauthorrefmark{1},
%Homer Simpson\IEEEauthorrefmark{2},
%James Kirk\IEEEauthorrefmark{3}, 
%Montgomery Scott\IEEEauthorrefmark{3} and
%Eldon Tyrell\IEEEauthorrefmark{4}}
%\IEEEauthorblockA{\IEEEauthorrefmark{1}School of Electrical and Computer Engineering\\
%Georgia Institute of Technology,
%Atlanta, Georgia 30332--0250\\ Email: see http://www.michaelshell.org/contact.html}
%\IEEEauthorblockA{\IEEEauthorrefmark{2}Twentieth Century Fox, Springfield, USA\\
%Email: homer@thesimpsons.com}
%\IEEEauthorblockA{\IEEEauthorrefmark{3}Starfleet Academy, San Francisco, California 96678-2391\\
%Telephone: (800) 555--1212, Fax: (888) 555--1212}
%\IEEEauthorblockA{\IEEEauthorrefmark{4}Tyrell Inc., 123 Replicant Street, Los Angeles, California 90210--4321}}

% use for special paper notices
%\IEEEspecialpapernot\usepackage[usenames, dvipsnames]{color}ice{(Invited Paper)}

% add copyright notice
\IEEEoverridecommandlockouts
\IEEEpubid{\makebox[\columnwidth]{\copyright2018 IEEE \hfill} \hspace{\columnsep}\makebox[\columnwidth]{ }}

% make the title area
\maketitle

% add copyright notice
\IEEEpubidadjcol

% As a general rule, do not put math, special symbols or citations
% in the abstract
%\textcolor{red}{R1: Abstract size reduction\\}
\begin{abstract}
Stringent latency requirements in advanced Internet of Things (IoT) applications as well as an increased load on cloud data centers have prompted a move towards a more decentralized approach, bringing storage and processing of IoT data closer to the end-devices through the deployment of multi-purpose IoT gateways. However, the resource constrained nature and diversity of these gateways pose a challenge in developing applications that can be deployed widely. This challenge can be overcome with containerization, a form of lightweight virtualization, bringing support for a wide range of hardware architectures and operating system agnostic deployment of applications on IoT gateways. This paper discusses the architectural aspects of containerization, and studies the suitability of available containerization tools for multi-container deployment in the context of IoT gateways. We present containerization in the context of AGILE, a multi-container and micro-service based open source framework for IoT gateways, developed as part of a Horizon 2020 project. Our study of containerized services to perform common gateway functions like device discovery, data management and cloud integration among others, reveal the advantages of having a containerized environment for IoT gateways with regard to use of base image hierarchies and image layering for in-container and cross-container performance optimizations. We illustrate these results in a set of benchmark experiments in this paper.
\end{abstract}


% For peer review papers, you can put extra information on the cover
% page as needed:
% \ifCLASSOPTIONpeerreview
% \begin{center} \bfseries EDICS Category: 3-BBND \end{center}
% \fi
%
% For peerreview papers, this IEEEtran command inserts a page break and
% creates the second title. It will be ignored for other modes.
\IEEEpeerreviewmaketitle

\section{Introduction}
% no \IEEEPARstart
Over the recent years, the concept of Internet of Things has gradually evolved from a paradigm to create a network of objects connected to the Internet to an interconnected network of data producers and data consumers. Regular day-to-day objects equipped with sensors act as data producers generating data sensed from the surrounding environment while software applications as well as end-devices equipped with actuators act as data consumers performing actions based on the data gathered. This notion of IoT has led to its application to various domains like health-care, autonomous transport, smart cities, among others which leverage the data generated from end devices to gain meaningful insights on the generated data. Due to the resource-constrained nature of these IoT end-devices, the scope of storage and data processing on these devices is limited. Thus, the data processing and business analytics are performed on cloud platforms and software services running on the cloud. However, with the number of connected devices predicted to grow up to 75 billion by 2025 \cite{statportal}, the data processing and storage architecture based solely on the cloud is facing a few challenges. 

\par The majority of IoT applications are heavily dependent on cloud storage and processing which affects the end-to-end latency and results in inefficient utilization of resources. These challenges, together with privacy and security considerations, have prompted a move away from the centralized architecture of storage and processing on the cloud to bring the processing and storage closer to the end-devices with the edge computing model. The edge computing model leverages a set of devices in the architecture between the end devices and the cloud. These devices can be legacy devices present in the network with storage and processing capabilities \cite{bonomi2012fog} or dedicated devices deployed to serve this purpose like IoT gateways and cloudlet devices\cite{verbelen2012cloudlets}. 

\par The presence of the edge computing layer aids offloading of data processing and validation to the edge layer. Moreover, it facilitates implementation of collaborative computing among end-devices as well as implementation of device and data management policies. However, the edge layer devices are not usually resource enriched; thus following the cloud oriented approach of hypervisor-based virtualization can prove to be cumbersome on these devices. Moreover, the edge layer devices are heterogeneous in terms of their hardware specifications, processor architecture and operating systems running on the devices. Thus, light-weight virtualization in the form of containerization offers a suitable solution to the concerns addressed above. Containerization allows virtualization at the OS-level, leveraging the kernel of the operating system to offer isolated user-spaces to each container. Thus, containerization facilitates the implementation of a microservice architecture, where each functionality on the edge device is developed as a service running inside each container. Containerization offers the flexibility to develop different services in different programming languages and communication among the containers using well-defined APIs.

\par In this paper, we present AGILE, an open-source framework for IoT gateways offering services including device and protocol management, data storage, security and access control. AGILE is designed based on a microservice architecture with each of the services above deployed in separate containers. Such containerization provides multiple advantages, but comes with a performance overhead. We use AGILE as a case-study to observe the overhead associated with containerization over a conventional approach and discuss improvements achieved by applying techniques like cross-container optimization and in-container optimization.

\par The rest of the paper is structured as follows. In section~\ref{sec:related}, we discuss about the state of the art in this research area highlighting the work on the performance measurements on containerized approaches as well as work on containerization in the IoT context. In section~\ref{sec:agile}, we discuss the services offered by the AGILE framework and the corresponding architecture. Section~\ref{sec:containerization} highlights the different approaches for optimizing the performance of containers for edge layer devices. In section V, we present our observations from the performance tests carried out on AGILE, concluding our work in section VI.

\begin{figure*}
\includegraphics[width=\textwidth]{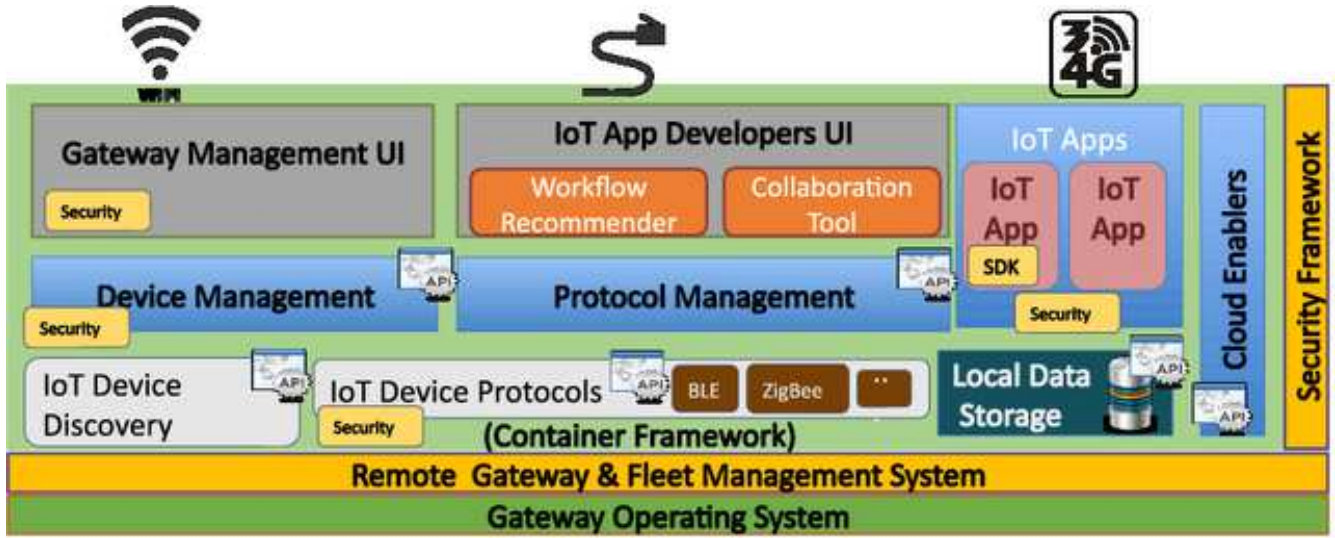}
\caption{Overview of the AGILE modular architecture}
\label{fig:agile_arch}
\end{figure*}

\section{Related Work}
%\textcolor{red}{R2: refer to work showing RPi is actually available as resources on the edge}
\label{sec:related}
The growth in both the number of IoT devices and IoT applications in a wide array of domains has brought about new requirements to the IoT ecosystem which include location awareness, geo-distribution of processing nodes and low latency in device-cloud communication. This has led to a burden on the traditional resources for IoT including networking, storage and processing resources. Consequently, considerable amount of literature has been published on the use of Single Board Computers (SBCs) like Raspberry Pi as an intermediate processing layer \cite{pahl2016container}\cite{chang2014bringing} and an enabler for various IoT applications\cite{senthilkumar2014embedded}\cite{morabito2016enabling}. Moreover, the suitability of networking resource virtualization for IoT including Software Defined Networks (SDNs) and Network Virtualization among others \cite{bizanis2016sdn}\cite{el2015software} have also been discussed in existing work. Our study on the existing literature focuses on virtualization of storage and processing resources at the OS-level in the form of containerization.  We study two aspects of containerization in the IoT context, first, the existing studies on the suitability and performance of containerization on resource-constrained devices and second, the application of containerization to different use-cases of IoT.

\subsection{Suitability of Containerization}
Previous studies have focused on the tradeoffs in applying hypervisor based virtualization and lightweight containerization to edge devices. The authors of \cite{pahl2015containers} illustrate the advantages of containerization over hypervisor based hardware virtualization in terms of size of the resources, flexibility and portability. The authors state that hypervisor based virtualization is more suited for Infrastructure-as-a-Service on the cloud than containerization, which offers a portable runtime, easier deployability on multiple servers and interconnectivity among containers. These advantages, on the other hand, make containerization more suitable for the edge layer in a Platform-as-a-Service scenario \cite{pahl2015containerization}. Pahl et. al \cite{pahl2016container} leverages the resources of Raspberry Pi devices to further build a cluster of containers running on multiple devices in the PaaS context. The cluster is designed to perform computationally intensive tasks including cluster and data management, overcoming the resource-constrained nature of each device. 

\par A significant amount of research is aimed at conducting performance tests to analyze the behavior of edge devices with the implementation of containerization. The authors of \cite{ramalho2016virtualization} study the performance of VM based virtualization and containerization against a native approach in terms of CPU cycles, disk throughput and network throughput. The results show that containerization outperforms VM based virtualization for memory I/O, network throughput and CPU cycles. Morabito et. al \cite{morabito2017evaluating} studies the processing resource and power consumption for performing different tasks on Raspberry Pi for wired and wireless connectivity. These tasks include running a containerized CoAP server for processing data, as well as for performing sensing actuation and video analytics. The author of \cite{krylovskiy2015internet} performs benchmark tests on Raspberry Pi B+ and Raspberry Pi model B in terms of Disk and Network I/O for native and containerized approaches. While the overhead is very high for the Raspberry Pi B+, significant performance improvements are observed for Raspberry Pi 2.  
\par In the above studies, the benchmarks clearly show that containerization is feasible on System on Chip (SOC) devices like Raspberry Pi 2 and it is more optimized than using VM based virtualization. However, there is a lack of studies on how the process of containerization itself can be optimized and benchmarks tests for the same. Existing literature shows approaches to deploy containers in a cluster of devices, however, there exists a gap in terms of multi-container deployment and optimization on a single device.

\subsection{Containerization in IoT use cases}
Several articles in existing literature present applications of IoT based on containerization in different use cases. The authors of \cite{rufino2017orchestration} demonstrate a distributed and layered architecture for Industrial IoT based applications with deployment of docker containers on end devices, the gateway and the cloud. Chesov et. al \cite{chesov2016containerized} present a multi-tier approach to containerize different functionalities in the smart cities context like data aggregation, business analytics and user interaction with data and deploy the containers on the cloud. Kovatsch et. al simplify programmability of IoT applications by exposing scripts and configurations using a RESTful CoAP interface from a deployed container \cite{kovatsch2012actinium}.

\par The existing literature applies containerization to different use-cases and specific areas of IoT. However, there is a lack of a framework based on containerization which can be applicable to multiple use cases and applications of IoT. We try to address this gap with AGILE, to build a modular framework based on containerization to offer services abstracted from various use cases and applications of IoT.

\section{AGILE Framework}
\label{sec:agile}
The AGILE gateway, which stands for an Adaptive \& Modular Gateway for the IoT, was conceived to design and implement a truly modular gateway in terms of hardware and software and to be adaptive to handle various types of devices, networking interfaces, communication protocols, and use cases. 

\subsection{AGILE Microservices}

AGILE is aimed at developing an open source software and hardware framework for IoT development. The hardware framework involves development of two separate versions of the gateway, a maker's version supporting fast prototyping based on the Raspberry Pi 3 board while the industrial version is being developed for use cases requiring rugged hardware and being ready for production. In the following subsection, we elaborate the software framework for AGILE as illustrated in Fig.~\ref{fig:agile_arch}.

\subsubsection{Device Management}
The device management handles addition and removal of new devices to the gateway. The device manager supports a set of devices which offer interfaces like reading and writing to the device as well as execution of methods offered by the device. The devices communicate using one or more protocols supported by the protocol manager.

\subsubsection{Protocol Management}
The protocol manager offers interfaces to add and remove protocols as well as support for implementations of underlying methods for each protocol including device discovery. The protocol manager supports a set of protocols which offer interfaces like read, write, connect and disconnect with the devices implementing the protocol.

\subsubsection{Local Data Storage}
The data storage component provides timeseries based storage for data generated by IoT sensors, with support for retention policies and encryption. 

\subsubsection{Gateway Management UI}
The gateway management UI allows the user to manage and access functionalities on the gateway, to start and stop other services like device discovery as well as access to data and visualization of the stored data. 

\subsubsection{IoT App Developers UI and SDK}
The IoT App Developers UI is aimed at offering a user-friendly graphical interface to create application logic by wiring together AGILE specific and generic nodes in an application workflow. The applications can e.g. include data collection and storage, rules applicable on the data to implement sensing-actuation use-cases, as well as analytics on sensor data or on the data stored on the gateway. A separate software development kit (SDK) facilitates development of IoT applications written direcetly in JavaScript, while apps written in other languages can also directly use the APIs provided by gateway microservices.

\subsubsection{Cloud Enablers}
The AGILE framework supports multiple cloud providers including Xively and Google Drive to process and push the data collected from the device and store them on these cloud platforms. This allows seamless end-to-end connectivity from the peripheral devices up to the cloud platforms.

\subsection{Containerized Architecture for AGILE}
The software framework for AGILE is designed using a microservice architecture. The rationale behind following the microservice oriented approach are the following: (i) The services offered are split into components which can interact with each other over required and provided interfaces, (ii) Using this approach, the components are easily scalable and adaptable to changing requirements in the system. 

The implementation of the software framework is achieved by containerizing the offered microservices. The containerization engine we have used for AGILE is Docker due to its wide-scale adoption, documentation and support for multiple architectures. The functionalities mentioned in the previous subsection are implemented individually in different Docker containers. %\textcolor{red}{R3 Why are they implemented in Docker containers? Isn't it already mentioned when we discuss related work? Otherwise we can write about isolation for each component, easier updates and modularity} 
The framework is language-agnostic, allowing development in any language, and thus, a wider choice of open-source code to be reused. Moreover, software dependency conflicts are easy to overcome since each containerized service has its own file system namespace.

The core functionality of the gateway which includes device management and protocol management is containerized exposing the interfaces for the HTTP REST API endpoints. The protocols are implemented in individual containers which include ZigBee and BLE. These core containers communicate with each other over the DBus which is implemented in a containerized form as well. 

The Gateway management UI is implemented using OS.js, a JavaScript based Web Desktop platform, in a containerized environment. This container depends on agile-core to offer access to the other microservices. The developers UI leverages the Node-RED tool deployed in a separate container. The cloud integration modules are provided as nodes inside the Node-RED tool. The number of containers used for implementing incremental versions of the AGILE stack is illustrated in Figure 2; which acts as a premise for our experimental setup.

\begin{figure}
\centering
\includegraphics[width=\columnwidth]{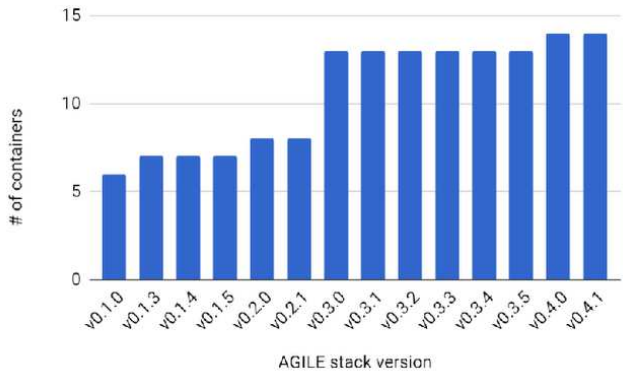}
\caption{Number of containers for each AGILE stack version} %\textcolor{red}{R2 doubts the results are not found through experimentation, although this is an experimental setup figure not depicting results}}
\label{fig_sim}
\end{figure}

\section{Containerization and IoT}
\label{sec:containerization}
Containerization of the micro-services provides several advantages in the design, development, deployment, security, and management of the gateways, however it also brings non-negligible overhead in other areas, most notably in resource consumption.

From the design perspective, the containerized architecture maps well to the micro-service principle, providing clear boundaries between individual services. Containerization also allows fine-grained control over access to system resources as well as restricted interactions between containers based on access policies. For example, in case of AGILE, only protocol adapters dealing with network interfaces are allowed access to system devices specifically only to those interfaces they require. Containerization facilitates version management of individual components, while also simplifying the whole micro-service composition. Finally, we should note that due to the large number of Edge/GW class hardware platforms, support of Linux distributions and package repositories is often lagging behind their server and desktop counterparts. Containerization overcomes this by requiring only an up-to date kernel and the container framework to be installed on the host. The rest is containerized and available on all CPU compatible platforms.  

Containerization can also simplify the development process if the build process is also containerized. In this case, tools in the host file system are not involved in the build process, ensuring that the service is built from source in an entirely reproducible way. 

However, there is considerable overhead involved in containerization, especially if we consider a multi-container installation with dozens of containers deployed on a single gateway class machine. Most prominent is the overhead in the overall size of deployment, which translates into an overhead in cost since higher reliability in the form of Embedded MultiMediaCard (eMMC) memory is more expensive than Secure Digital (SD) cards. Due to the possibility of limited and intermittent connectivity, download or update times affect the performance of the system. Due to the resource constrained nature of the gateways in comparison to the cloud, high resource utilization and build times for containers add to the overhead.

As mentioned earlier AGILE relies on Docker for containerization. In a Docker based multi-container environment each micro-service has its own file system image generated using a ``Dockerfile'': the shell script like ``recipe'' to build, install, and run the service. To allow sharing content between containers and to speed up generation of images, Docker uses layered images where each layer contains a set of files adding, overwriting, or deleting files from the union of previous layers. The image generation starts from an image referred to as the ``base image'', followed by addition of build dependencies, build tools, build artifacts, runtime dependencies, and language runtimes. The overall size for all the images of all micro-services can be significant.

To reduce the overall size of the distribution and to address the aforementioned overhead we define a novel taxonomy to propose the following optimization techniques.

\begin{figure}[t]
\includegraphics[width=\columnwidth]{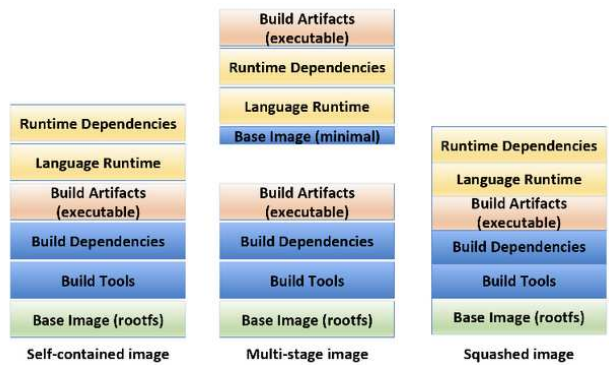}
\caption{Layered image stack for in-container optimization}
\label{layering-diagram-in-container}
\end{figure}

\subsection{In-container image layering optimizations}
In container image layering optimization is aimed at reducing 
overhead during the generation and deployment of individual micro-services. Techniques illustrated on Fig.~\ref{layering-diagram-in-container} are:
\subsubsection{self-contained Dockerfile} In this case, which serves as our baseline for the following optimizations, each of the above mentioned steps in the process of generating the final image from the base image generates one layer or more in the layered file-system of the container depending on the complexity of each step. 
\subsubsection{multi-stage Dockerfile} In case of the multi-stage build, first a dockerized ``build image'' is created comprising of the base image, build dependencies and build steps. Leveraging the build image, a special ``deployment image'' is created containing only the runtime environment and the actual executables\footnote{While previously multi-stage build required external tools, support was included in docker with version 17.05-ce. With this we have also migrated our containers to use the new framework.}.
\subsubsection{image squashing} The image squashing technique creates a single-layered or monolithic image from a multi-layered image. In a multi-layered image, the generation of a new layer makes all previous layers immutable. Thus, if a file is modified or deleted, the old content is residual, increasing overall image size. This overhead of larger image sizes is mitigated by the image squashing technique.

\begin{figure}[t]
\includegraphics[width=\columnwidth]{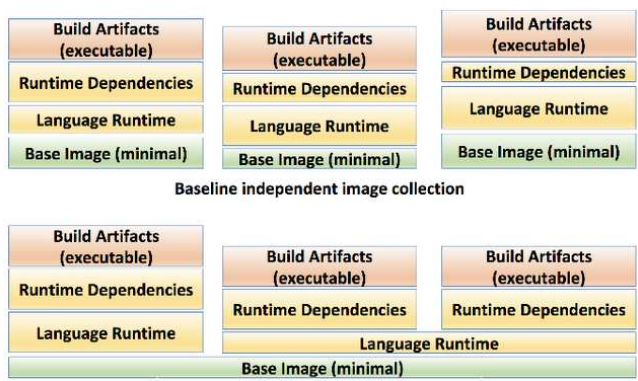}
\caption{Layered image stack for cross-container optimization}
		\label{layering-diagram-cross-container}
\end{figure}

\subsection{Cross-container optimizations}
When multiple containers are deployed on the same device, a further optimization is possible as shown on Fig.~\ref{layering-diagram-cross-container}: containers from the set of images can rely on common layers.
\subsubsection{baseline image collection} Our baseline is a multi-container setup where each image is generated by developers choosing their base images independently. In practice, this can result in a lack of common layers among the images, and the overall distribution size becomes the sum of individual images.
\subsubsection{base image hierarchy based optimization} To overcome the previous problem, we have introduced images based on a hierarchy of base images with the objective of maximizing the number of shared base layers. This hierarchy\footnote{We rely on and contribute to the open source Resin.io image hierarchy.} of base images is built as follows: firstly, a series of lean CPU architecture specific images are created. Secondly, based on each of these images, a series of Linux distribution specific images are created. The significance of this second layer is to provide broad support for installation of both build and runtime dependencies. In the third layer of the hierarchy, images of the previous layer are leveraged to generate images for every relevant language build environment and runtime. By forcing the selection of images from this hierarchy, we achieve two important goals: (i) the number of layers that are shared between our deployed images is maximized, for example a Java Runtime Environment (JRE) will only be deployed once, and not in all images running Java code. (ii) Since all base images in the hierarchy exist for all CPU architectures, supporting multiple architectures becomes straightforward by generating CPU specific versions of our own images.

\section{Observation}
In this section, we present our study of the performance of our multi-container microservice based deployment on the AGILE gateway. We study the performance improvements in terms of the container sizes and container download/installation times.

\subsection{Experimental setup}
For the experiments, we have used the makers' version of the gateway which consists of an ARM Cortex A53 processor clocked at 1.2 GHz, 1 GB of RAM and a 32 GB Samsung EVO Plus Class 10 SD card. The installed operating system on the device was Raspbian Jessie, and it was connected to the Internet with a dedicated 50/10 Mbps FTTC/VDSL2 line. The following containers comprising the AGILE stack were used for the experiment. The updates to the overall stack were pushed and released monthly while individual containers were updated biweekly on average.
\subsubsection{agile-core}: The agile-core container is built from a ZuluJDK base image and is dependent on the following containers, (i) agile-dbus, (ii) agile-devicemanager, (iii) agile-protocolmanager and (iv) agile-devicefactory. If the containers required by agile-core are already built and available on the device, they are reused, or they are built as well during the build process. 

\subsubsection{agile-ble}: The agile-ble container is also built from a ZuluJDK base image and depends on the agile-core docker image. The agile-ble container registers a dbus object which can be leveraged by the agile-protocolmanager. The dbus object offers interfaces for methods supported by the BLE protocol including connection, discovery, read and write methods.

\subsubsection{agile-nodered}: The agile-nodered container is built from a NodeJS base image and does not depend on other containers. This container runs the Node-Red application inside the container and exposes an endpoint to access the application. The Node-Red running inside the container also offers access to multiple other custom Node-Red nodes developed for the AGILE gateway which includes recommendation, cloud integration and device interfacing nodes.

\subsection{Tests}
\noindent %\textcolor{red}{R2: 1. Mention frequency of updates\\ R2 2. Elaborate why the base containers can't be deployed on the gateway and consider only incremental udpates, which is a valid point and something we actually used to do with AGILE, start with a specific version of the stack and then update it to the latest. \\R2 3. Is 2 GB of space really required for computation? elaborate}
The following tests are conducted to address two key issues for gateway devices mentioned in the state of the art. First, we study the sizes of the selected images as a function of in-container image optimization techniques, assessing their effectiveness. Second, we evaluate cross-container optimization by looking at the size and consequently download time of the AGILE stack. Time measurements are repeated ten times and average values are presented.
%Finally, build times on constrained gateways can be significantly time consuming. Although some companies start to offer native build farms as a service, build times still provide interesting insight into the cost of containerization.

 \begin{table}[htbp]
 \centering
 \caption{Image sizes obtained by in-container optimization}
 \label{tab:image-size}
 \begin{tabular}{|l|l|l|l|}
  \hline
  Image & Baseline  & Multi-stage  & Squashed    \\
  \hline
  agile-ble & 801MB & 372MB & 361MB \\
  \hline
  agile-nodered & 613MB & 294MB & 291MB \\
  \hline
  agile-core & 789MB & 348MB & 342MB \\
  \hline
%  AGILE stack &  &  & \\
%  \hline
 \end{tabular}
 \end{table}

Table \ref{tab:image-size} shows the effect of in-container optimization techniques on images of the three selected components. Note that image sizes contain base layers, language runtimes for different languages (Java or JavaScript) and dependencies, hence the relatively large image sizes. Multi-stage build optimization reduces image sizes to as low as 45\%, in case of agile-core, of the baseline. Squashing, if applied on top of the already optimized multi-stage image, improves image size only marginally.

\begin{figure}[t]
\includegraphics[width=\columnwidth]{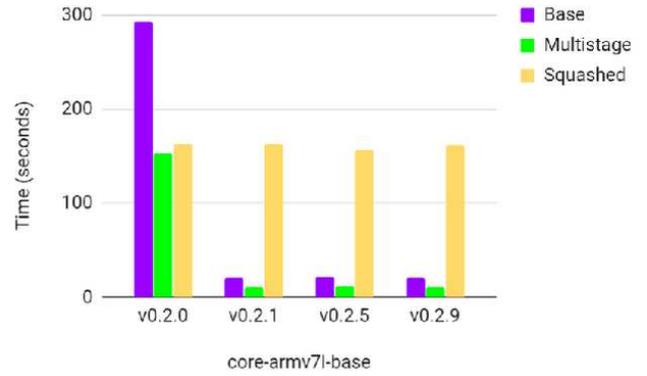}
\caption{Download times for in-container optimization techniques}
\label{fig:image-dl}
\end{figure}

On the contrary, as we show next, squashing can be counterproductive for updates. Fig.~\ref{fig:image-dl} shows the download times for images built using in-container optimization. Although in reality we have introduced various optimizations gradually as the stack and its components evolved in the recent years, for this experiment we have regenerated each version of the components in a baseline, a multi-stage, and a squashed version. First, the generated images are pushed to repositories on Docker Hub with incremental tags for changing versions. These images are then iteratively downloaded in the following manners: (i) \emph{absolute}, where previous versions of an image are removed before download, and (ii) \emph{incremental}, emulating software update, where previous image versions are kept to download just the updated layers.

The first column of figure \ref{fig:image-dl} shows the absolute download times of a specific version of agile-core, considering the three aforementioned techniques. Download times are almost proportional to image sizes, although it is interesting to note that the squashed image, even if smaller, downloads slightly slower than the multi-stage one. This is due to the way image download and decompression is parallelized in Docker. Subsequent columns show incremental download performance, where multi-stage, even baseline outperforms squashing. In fact, squashing forces the whole image content to be downloaded again, while multi-stage allows small changes to transform into the update of only a few small layers.

\begin{figure}[t]
\includegraphics[width=\columnwidth]{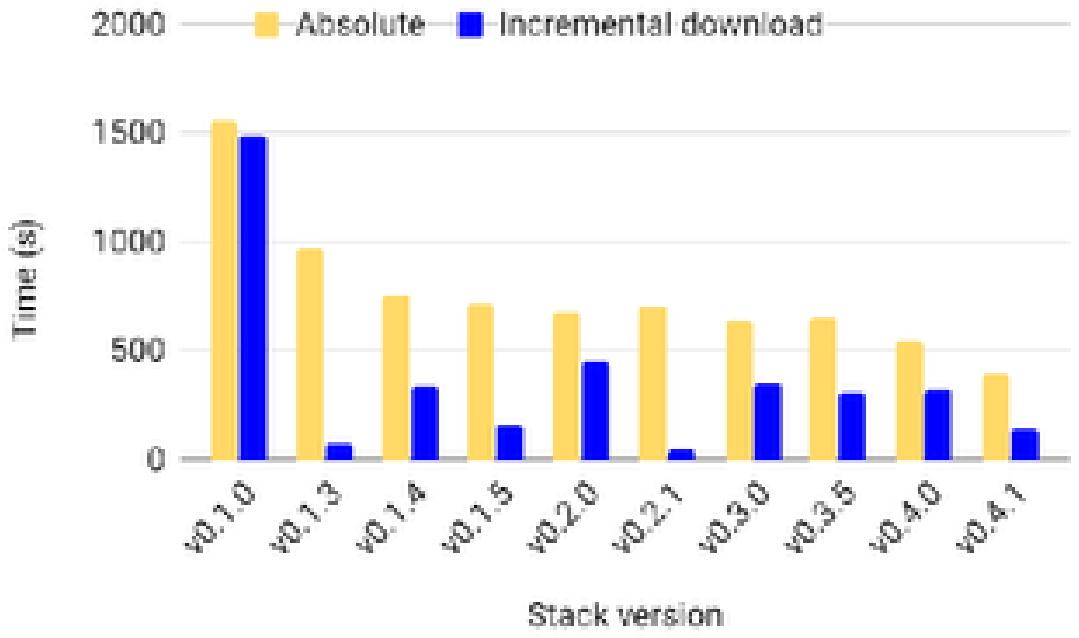}
\includegraphics[width=\columnwidth]{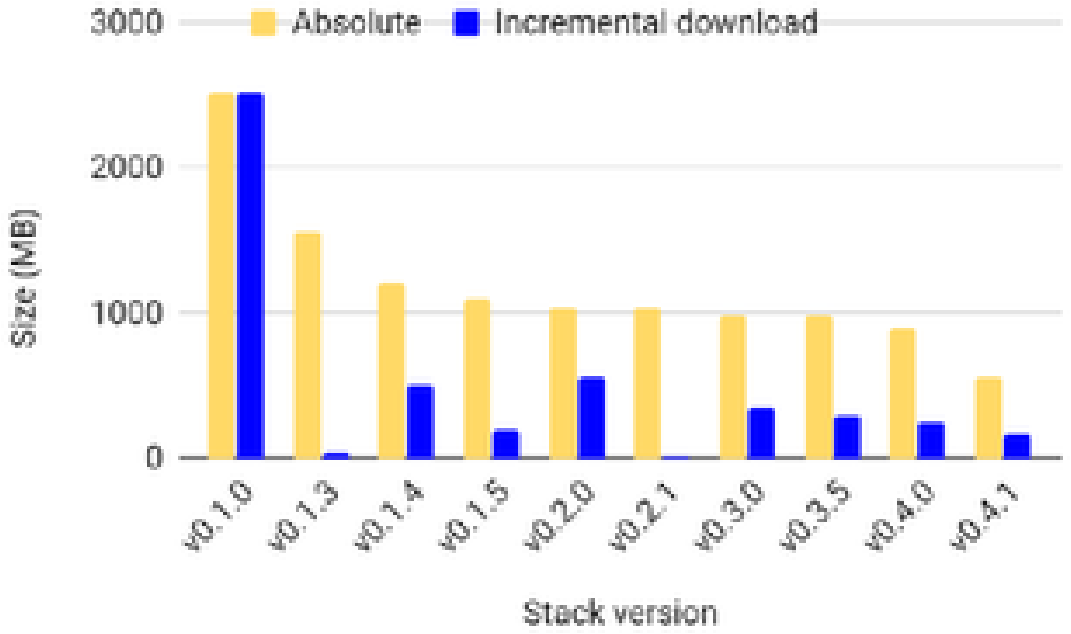}
\caption{Download times and data amounts for cross-container optimization techniques}
\label{download-times}
\end{figure}

Finally, we consider multiple containers on the same gateway, i.e. the deployment of the AGILE stack. To simplify discussion and avoid distortion from functionalities added during the evolution of the stack, we restrict deployment to the subset of images discussed before (agile-core, agile-ble, agile-nodered). Fig.~\ref{download-times} shows both the download time and the data amount as the stack evolved. In the initial versions, neither the base images were chosen considering common layers, nor in-container optimizations were used. Hence, the download amount was similar to the sum of baseline image sizes in Table~\ref{tab:image-size}. The difference is due to the way data was obtained: in this experiment we use the original images forming a given version of the stack, while Table~\ref{tab:image-size} contains the size of the newest version of each component regenerated using a given optimization technique. Incremental times and size also shows large values for v0.1.0, but only because this is the first stack deployed on a clean system. In the subsequent versions, absolute download times are reduced due to both cross-container and in-container optimizations. More specifically, multi-stage was introduced in v0.1.3 for agile-core and agile-ble, while only in v0.4.1 for agile-nodered. Reduction in the overall size can clearly be seen in these cases. The use of the base image hierarchy, instead, was introduced in v0.1.4 and v0.2.0, respectively. Since this involved a change in the base images, incremental downloads increased for the specific release, but the overall effect to stack size is positive.

\section{Conclusion}
%The conclusion goes here.
This paper summarizes the current state of the art on containerization techniques and suitability of containerization in the context of IoT systems. The existing gap in the current literature in terms of multi-container deployments on edge layer devices is addressed by illustrating a microservice architecture based container deployment and defining optimization techniques to improve the gateway performance.

The cross-container optimization technique proposed facilitates the reduction in build times for images and removes the redundancy among base images used in the stack. On the other hand, in-container optimization reduces the overall size of the AGILE stack and thus reduces download times for the images as well.

In future work, we will perform further measurements on multiple devices and device architectures to improve our proposed optimization techniques. We would also investigate optimization of pushing delta updates to the devices while minimizing the build and download times using the current study as a reference.

% conference papers do not normally have an appendix

% use section* for acknowledgment
\section*{Acknowledgment}

This work was funded by the European Commission under the H2020 AGILE (Adaptive Gateways for dIverse muLtiple Environments) project; project ID: 688088. The authors would like to thank all participants of the AGILE project.

%The authors would like to thank...

% trigger a \newpage just before the given reference
% number - used to balance the columns on the last page
% adjust value as needed - may need to be readjusted if
% the document is modified later
%\IEEEtriggeratref{8}
% The "triggered" command can be changed if desired:
%\IEEEtriggercmd{\enlargethispage{-5in}}

% references section

% can use a bibliography generated by BibTeX as a .bbl file
% BibTeX documentation can be easily obtained at:
% http://mirror.ctan.org/biblio/bibtex/contrib/doc/
% The IEEEtran BibTeX style support page is at:
% http://www.michaelshell.org/tex/ieeetran/bibtex/
\bibliographystyle{IEEEtran}
% argument is your BibTeX string definitions and bibliography database(s)
\bibliography{references}

% Generated by IEEEtran.bst, version: 1.14 (2015/08/26)
\begin{thebibliography}{10}
\providecommand{\url}[1]{#1}
\csname url@samestyle\endcsname
\providecommand{\newblock}{\relax}
\providecommand{\bibinfo}[2]{#2}
\providecommand{\BIBentrySTDinterwordspacing}{\spaceskip=0pt\relax}
\providecommand{\BIBentryALTinterwordstretchfactor}{4}
\providecommand{\BIBentryALTinterwordspacing}{\spaceskip=\fontdimen2\font plus
\BIBentryALTinterwordstretchfactor\fontdimen3\font minus
  \fontdimen4\font\relax}
\providecommand{\BIBforeignlanguage}[2]{{%
\expandafter\ifx\csname l@#1\endcsname\relax
\typeout{** WARNING: IEEEtran.bst: No hyphenation pattern has been}%
\typeout{** loaded for the language `#1'. Using the pattern for}%
\typeout{** the default language instead.}%
\else
\language=\csname l@#1\endcsname
\fi
#2}}
\providecommand{\BIBdecl}{\relax}
\BIBdecl

\bibitem{statportal}
``{Internet of Things (IoT) connected devices installed base worldwide from
  2015 to 2025 (in billions)},''
  \url{https://www.statista.com/statistics/471264/iot-number-of-connected-dev
  ices-worldwide/}.

\bibitem{bonomi2012fog}
F.~Bonomi, R.~Milito, J.~Zhu, and S.~Addepalli, ``{Fog Computing and Its Role
  in the Internet of Things},'' in \emph{Proceedings of the first edition of
  the MCC workshop on Mobile cloud computing}.\hskip 1em plus 0.5em minus
  0.4em\relax ACM, 2012, pp. 13--16.

\bibitem{verbelen2012cloudlets}
T.~Verbelen, P.~Simoens, F.~De~Turck, and B.~Dhoedt, ``Cloudlets: Bringing the
  cloud to the mobile user,'' in \emph{Proceedings of the third ACM workshop on
  Mobile cloud computing and services}.\hskip 1em plus 0.5em minus 0.4em\relax
  ACM, 2012, pp. 29--36.

\bibitem{pahl2016container}
C.~Pahl, S.~Helmer, L.~Miori, J.~Sanin, and B.~Lee, ``{A container-based edge
  cloud PaaS architecture based on Raspberry Pi clusters},'' in \emph{Future
  Internet of Things and Cloud Workshops (FiCloudW), IEEE International
  Conference on}.\hskip 1em plus 0.5em minus 0.4em\relax IEEE, 2016, pp.
  117--124.

\bibitem{chang2014bringing}
H.~Chang, A.~Hari, S.~Mukherjee, and T.~Lakshman, ``Bringing the cloud to the
  edge,'' in \emph{Computer Communications Workshops (INFOCOM WKSHPS), 2014
  IEEE Conference on}.\hskip 1em plus 0.5em minus 0.4em\relax IEEE, 2014, pp.
  346--351.

\bibitem{senthilkumar2014embedded}
G.~Senthilkumar, K.~Gopalakrishnan, and V.~S. Kumar, ``Embedded image capturing
  system using raspberry pi system,'' \emph{International Journal of Emerging
  Trends \& Technology in Computer Science}, vol.~3, no.~2, pp. 213--215, 2014.

\bibitem{morabito2016enabling}
R.~Morabito, R.~Petrolo, V.~Loscr{\'\i}, and N.~Mitton, ``Enabling a
  lightweight edge gateway-as-a-service for the internet of things,'' in
  \emph{Network of the Future (NOF), 2016 7th International Conference on
  the}.\hskip 1em plus 0.5em minus 0.4em\relax IEEE, 2016, pp. 1--5.

\bibitem{bizanis2016sdn}
N.~Bizanis and F.~A. Kuipers, ``{SDN and virtualization solutions for the
  Internet of Things: A survey},'' \emph{IEEE Access}, vol.~4, pp. 5591--5606,
  2016.

\bibitem{el2015software}
A.~El-Mougy, M.~Ibnkahla, and L.~Hegazy, ``{Software-defined wireless network
  architectures for the Internet-of-Things},'' in \emph{Local Computer Networks
  Conference Workshops (LCN Workshops), 2015 IEEE 40th}.\hskip 1em plus 0.5em
  minus 0.4em\relax IEEE, 2015, pp. 804--811.

\bibitem{pahl2015containers}
C.~Pahl and B.~Lee, ``Containers and clusters for edge cloud architectures--a
  technology review,'' in \emph{Future Internet of Things and Cloud (FiCloud),
  2015 3rd International Conference on}.\hskip 1em plus 0.5em minus 0.4em\relax
  IEEE, 2015, pp. 379--386.

\bibitem{pahl2015containerization}
C.~Pahl, ``{Containerization and the PaaS Cloud},'' \emph{IEEE Cloud
  Computing}, vol.~2, no.~3, pp. 24--31, 2015.

\bibitem{ramalho2016virtualization}
F.~Ramalho and A.~Neto, ``Virtualization at the network edge: A performance
  comparison,'' in \emph{World of Wireless, Mobile and Multimedia Networks
  (WoWMoM), 2016 IEEE 17th International Symposium on A}.\hskip 1em plus 0.5em
  minus 0.4em\relax IEEE, 2016, pp. 1--6.

\bibitem{morabito2017evaluating}
R.~Morabito, I.~Farris, A.~Iera, and T.~Taleb, ``{Evaluating Performance of
  Containerized IoT Services for Clustered Devices at the Network Edge},''
  \emph{IEEE Internet of Things Journal}, 2017.

\bibitem{krylovskiy2015internet}
A.~Krylovskiy, ``{Internet of Things gateways meet linux containers:
  Performance evaluation and discussion},'' in \emph{Internet of Things
  (WF-IoT), 2015 IEEE 2nd World Forum on}.\hskip 1em plus 0.5em minus
  0.4em\relax IEEE, 2015, pp. 222--227.

\bibitem{rufino2017orchestration}
J.~Rufino, M.~Alam, J.~Ferreira, A.~Rehman, and K.~F. Tsang, ``{Orchestration
  of containerized microservices for IIoT using Docker},'' in \emph{Industrial
  Technology (ICIT), 2017 IEEE International Conference on}.\hskip 1em plus
  0.5em minus 0.4em\relax IEEE, 2017, pp. 1532--1536.

\bibitem{chesov2016containerized}
R.~Chesov, V.~Solovyev, Ð.~Khlamov, and Ð.~Prokofyev, ``Containerized cloud
  based technology for smart cities applications,'' \emph{Journal of
  Fundamental and Applied Sciences}, vol.~8, no.~3S, pp. 2638--2646, 2016.

\bibitem{kovatsch2012actinium}
M.~Kovatsch, M.~Lanter, and S.~Duquennoy, ``{Actinium: A RESTful runtime
  container for scriptable Internet of Things applications},'' in
  \emph{Internet of Things (IOT), 2012 3rd International Conference on
  the}.\hskip 1em plus 0.5em minus 0.4em\relax IEEE, 2012, pp. 135--142.

\end{thebibliography}
%
% <OR> manually copy in the resultant .bbl file
% set second argument of \begin to the number of references
% (used to reserve space for the reference number labels box)

% \section*{NOTES}
% \begin{itemize}
% \item Introduction
% \begin{itemize}
% \item IoT growth+processing on cloud = leading to greenhouse gases, bottleneck on DCNs, poor latency
% \item Deployment of Edge devices to address this issue
% \item However, resource constrained IoT gateways, unable to support full fledged VMs
% \item Hardware diversity, lack of standardization, not possible to write one app and deploy on multiple gateways 
% \item Containerization stack, how is it better than deploying VMs?
% \item Need for containerization, breakdown functionality into applications and containerize them
% \item Advantages: lightweight, secure, OS-agnostic
% \item Paper structure 
% \end{itemize}
% \end{itemize}
% % that's all folks
% \begin{itemize}
% \item Literature Review
% \begin{itemize}
% \item Previous works on Containerization for IoT
% \item Challenges for bringing Containerization to IoT gateways
% \item Problem statement \textcolor{red}{How to present the problem statement?}
% \end{itemize}
% \end{itemize}
\end{document}